\documentclass[twocolumn,journal]{IEEEtran}
\usepackage[T1]{fontenc}
\usepackage[latin9]{inputenc}
\usepackage{float}
\usepackage{amstext}
\usepackage{cite}
\usepackage{amsthm}
\usepackage{amssymb}
\usepackage{graphicx}
\usepackage{esint}
\usepackage[unicode=true,
 bookmarks=true,bookmarksnumbered=true,bookmarksopen=true,bookmarksopenlevel=1,
 breaklinks=false,pdfborder={0 0 0},pdfborderstyle={},backref=false,colorlinks=false]
 {hyperref}
\hypersetup{pdftitle={Your Title},
 pdfauthor={Your Name},
 pdfpagelayout=OneColumn, pdfnewwindow=true, pdfstartview=XYZ, plainpages=false}
\usepackage{breakurl}

\makeatletter

\floatstyle{ruled}
\newfloat{algorithm}{tbp}{loa}
\providecommand{\algorithmname}{Algorithm}
\floatname{algorithm}{\protect\algorithmname}

\let\oldforeign@language\foreign@language
\DeclareRobustCommand{\foreign@language}[1]{%
  \lowercase{\oldforeign@language{#1}}}
\theoremstyle{plain}
\newtheorem{thm}{\protect\theoremname}
\theoremstyle{plain}
\newtheorem{prop}[thm]{\protect\propositionname}

\usepackage[caption=false,font=footnotesize]{subfig}

\makeatother

\providecommand{\propositionname}{Proposition}
\providecommand{\theoremname}{Theorem}

\begin{document}

\title{Secure Beamforming and Ergodic Secrecy Rate Analysis for Amplify-and-Forward
Relay Networks with Wireless Powered Jammer}

\author{Omer Waqar,~\IEEEmembership{Member,~IEEE,} Hina Tabassum,~\IEEEmembership{Senior Member, IEEE}
and~Raviraj Adve,~\IEEEmembership{Fellow, IEEE}\thanks{O. Waqar is with the Department of Engineering \& Applied Science, Thompson Rivers University (TRU), Canada. e-mail: \protect\href{http://xxx@xxx.xxx}{owaqar@tru.ca}.}\thanks{H. Tabassum is with the Department of Electrical Engineering and Computer
Science, York University, Canada. e-mail: \protect\href{http://xxx@xxx.xxx}{hina@eecs.yorku.ca}.}\thanks{R. Adve is with the Department of Electrical \& Computer Engineering, University of Toronto, Canada.  e-mail: \protect\href{http://xxx@xxx.xxx}{rsadve@ece.utoronto.ca}.}}

\markboth{}{Your Name \MakeLowercase{\emph{et al.}}: Your Title}
\maketitle
\begin{abstract}
In this correspondence, we consider an amplify-and-forward relay network
in which relayed information is overheard by an eavesdropper. In order
to confound the eavesdropper, a wireless-powered jammer is also considered
which harvests energy from a multiple-antenna source. We proposed
a new secure beamforming scheme in which beamforming vector is a linear
combination of the energy beamforming (EB) and information beamforming
(IB) vectors. We also present a new closed-form solution for the proposed
beamforming vector which is shown to achieve a higher secrecy rate
as compared to the trivial EB and IB vectors. Moreover, a tight closed-form
approximation for the ergodic secrecy rate is also derived for the
asymptotic regime of a large number of antennas at the source. Finally,
numerical examples and simulations are provided which validate our
analytical results.
\end{abstract}


\IEEEpeerreviewmaketitle{}

\section{Introduction}

\IEEEPARstart{O}{wing} to the benefits of the relaying architectures
in increasing the network coverage and/or its performance, an overwhelming
interest for wireless relay networks has been exhibited by both industry
and academia over the past decade \cite{Hua2013}. Relays play a critical
role in establishing a communication between the transmitter and the
receiver particularly when a direct link between these two nodes does
not exist. However, due to the broadcast nature of the wireless medium,
in addition to the legitimate receiver, relayed information can also
be overheard by a malicious eavesdropper. 

In order to safeguard the confidential message from wiretapping, physical
layer security (PLS) is considered as a promising security mechanism
which eliminates the drawbacks of the conventional cryptographic techniques
\cite{Mukherjee2014}. The ultimate goal of PLS is to enhance the
secrecy rate which is defined as the rate difference between the legitimate
channel and the wiretap channel \cite{Chen2016}. The significance
of the secrecy rate lies in the fact that secure communications (i.e.,
eavesdropper is unable to decode a confidential message) is possible
only when this rate is positive.

To increase the secrecy rate, one of the most effective approaches
is to degrade the wiretap channel through controlled artificial noise
(AN), as proposed in \cite{Goel2008}. Following the work of \cite{Goel2008},
several strategies have been devised for wireless relay networks under
the umbrella of cooperative jamming, e.g., in \cite{Li2015, Park2013, Zhao2016}
(see \cite{Jameel2019} for a comprehensive list of works regarding
cooperative jamming). In \cite{Li2015}, Li \textit{et al.} considered
multiple multi-antenna relays and eavesdroppers where each relay participates
in cooperative beamforming while emitting AN. In \cite{Park2013},
jamming power allocation strategies have been investigated for an
amplify-and-forward (AF) relay network in which the destination transmits
an intended jamming signal via relay to confuse an eavesdropper. New
closed-form expressions for the ergodic achievable secrecy rate have
been derived in \cite{Zhao2016} with a multiple-antenna AF relay
considering three different secure transmission schemes. 

Despite the fact that in many wireless applications, the nodes are
energy-constrained and have limited battery-life, most of the works
including the papers mentioned above assumed that the nodes which
generate jamming signals either are equipped with batteries of infinite
capacities or these jammers are connected to a fixed power supply.
To this end, wireless energy harvesting through radio frequency (RF)
signals has emanated as a promising solution which facilitates to
provide uninterruptible and controlled power to the energy-constrained
jammers \cite{Xing2015}, \cite{Mobini2019} and references therein.
In contrast to the existing works, we propose a new secure beamforming
scheme with multiple antennas at the source and a wireless powered
jammer for an AF relay network. 

In particular, our main contributions are: \textbf{(i)} We present
a new closed-form solution for the beamforming vector which maximizes
the instantaneous secrecy capacity. This closed-form solution provides
direct insights into the impact of various system parameters on the
proposed beamforming scheme. Moreover, as our proposed secure beamforming
scheme is a linear combination of energy beamforming (EB) and information
beamforming (IB), it achieves a higher ergodic secrecy rate (ESR)
when compared to the relatively straightforward EB and IB schemes;
\textbf{(ii)} We provide a new closed-form approximation for the ESR
with a given beamforming vector and it is shown that the approximation
is tight for large number of antennas at the source. Therefore, the
closed-form approximation evaluates the ESR efficiently without resorting
to time-consuming simulations.

\textit{Notation}s: $||\mathbf{x}||$ and $\mathbf{x^{\dagger}}$
denote the Euclidean norm and conjugate-transpose of vector $\mathbf{x}$,
respectively. The norm of a complex number $z$ is denoted by |$z$|.
Moreover $\mathbb{E}(\cdot)$ denotes expectation operator. $\left[y\right]^{+}\triangleq\max\left(y,0\right)$
and $f_{X}\left(x\right)$ represents probability density function.
${\textstyle E}_{i}(\cdot)$, ${\textstyle E}_{1}(\cdot)$ are exponential
integral functions, $\ln\{\cdot\}$ is natural logarithmic function and
$G_{p,q}^{m,n}\left[\cdot\right]$ is Meijer-G function \cite{Prudnikov1992}.

\section{System Model and Signal-to-Noise Ratio}

We consider a five-node network which comprises of a source (\textbf{S}),
a trusted variable-gain relay (\textbf{R}), an external jammer (\textbf{J)},
legitimate destination (\textbf{D}) and an eavesdropper (\textbf{E}).
All nodes are equipped with a single-antenna except \textbf{S} which
has $N\left(N\geq1\right)$ antennas. Moreover, all nodes have access
to fixed power supplies except \textbf{J} which harvests energy using
RF signals originating from\textbf{ S}. Furthermore, it is assumed
that both \textbf{R} and \textbf{J} operate in half-duplex mode with
AF relaying protocol is adopted at \textbf{R}. Moreover, as in \cite{Hoang2017}
we consider that the AN from\textit{ }\textbf{J} is either nulled
out at \textbf{D} through cooperation or it does not reach to \textbf{D}
due to an obstacle between \textbf{J} and \textbf{D}. Similarly, the
direct link between \textbf{S} and\textbf{ D} (and \textbf{E}) does
not exist due to poor channel conditions. We consider block fading
channels i.e., all channels remain constant for a block time \textit{T}
and then change independently. The communication takes place using
two phases each of length \textit{T}/2. 

During phase 1, the signals received at\textit{ }\textbf{J}\textit{
$\left(y_{J}\right)$ }and \textbf{R} \textit{$\left(y_{R}\right)$}
are, respectively, given as follows \vspace{-0.2cm}

\begin{equation}
y_{J}=\sqrt{{\displaystyle \frac{P_{s}L_{c}}{d_{SJ}^{\alpha}}}}\mathbf{h}_{SJ}^{\dagger}\mathbf{w}x_{S}+n_{J},\label{eq:received_signal_Jammer}
\end{equation}
\vspace{-0.1cm}
\begin{equation}
y_{R}=\sqrt{{\displaystyle \frac{P_{s}L_{c}}{d_{SR}^{\alpha}}}}\mathbf{h}_{SR}^{\dagger}\mathbf{w}x_{S}+n_{R},\label{eq:received_signal_Relay}
\end{equation}
where $x_{S}$ is an information signal transmitted from \textbf{S}
with $\mathbb{E}\left\{ |x_{S}|^{2}\right\} =1$ and $P_{s}$ is the
transmit power of \textbf{S}. Moreover, $\mathbf{w}$ is $N\times1$
unit-norm beamforming vector. $d_{ij}$ denotes the distance between
the $i$-th and $j$-th node where $i$, $j$ $\in$ (\textbf{S}, \textbf{R},
\textbf{J}, \textbf{D} and \textbf{E}). $L_{c}$ and $\alpha$ represent
path-loss constant and a path-loss exponent, respectively. Furthermore,
$\mathbf{h}_{SJ}$ and $\mathbf{h}_{SR}$ are $N\times1$ channel
vectors for the \textbf{S} to\textit{ }\textbf{J} and \textbf{S} to
\textbf{R} links, respectively with mutually independent complex entries
following complex circularly symmetric Gaussian (CSG) distributions
of zero mean and unit variance. Moreover, $n_{i}$ denotes the additive
white Gaussian noise (AWGN) at $i$-th node following complex CSG
distribution with zero mean and variance $\sigma_{n}^{2}$ . 

During phase 2, the signal transmitted by \textbf{J} is $x_{J}=\sqrt{P_{J}}v$,
where $P_{J}$ is the transmit power of \textbf{J} and $v$ is a unit-power
AN signal. Using (\ref{eq:received_signal_Jammer}) and assuming that
the noise power is negligible for energy harvesting purposes, we can
write\vspace{-0.2cm}

\begin{equation}
x_{J}=\sqrt{\eta P_{s}L_{c}d_{SJ}^{{\scriptstyle -}\alpha}}|\mathbf{h}_{SJ}^{\dagger}\mathbf{w}|v,\label{eq:transmitted_signal_Jammer}
\end{equation}
where $\eta\,\in\,\left[0,1\right]$ represents an energy conversion
efficiency. Moreover, the signal transmitted by \textbf{R}\textit{
}is\textit{ $x_{R}=\beta y_{R}$, }where \textit{ }$\beta$ is an
amplification gain factor which satisfies $\mathbb{E}\left\{ |x_{R}|^{2}\right\} =P_{R}$.
Therefore, we can write \vspace{-0.3cm}

\begin{equation}
\beta^{2}={\displaystyle \frac{P_{R}}{P_{S}L_{c}d_{SR}^{-\alpha}|\mathbf{h}_{SR}^{\dagger}\mathbf{w}|^{2}+\sigma_{n}^{2}}},\label{eq:Beta_eq}
\end{equation}
where $P_{R}$ denotes the transmit power of \textbf{R}. The signals
received at \textbf{D }$\left(y_{D}\right)$ and \textbf{E} $\left(y_{E}\right)$
are , respectively, given as\vspace{-0.3cm}

\begin{equation}
y_{D}=\beta\sqrt{{\displaystyle \frac{P_{s}L_{c}^{2}}{\left(d_{SR}d_{RD}\right)^{\alpha}}}}\left(\mathbf{h}_{SR}^{\dagger}\mathbf{w}\right)h_{RD}x_{S}+{\displaystyle \frac{\beta h_{RD}n_{R}}{\sqrt{L_{c}d_{RD}^{\alpha}}}}+n_{D},\label{eq:received_signal_at_D}
\end{equation}
\vspace{-0.4in}

\begin{equation}
\begin{array}[b]{c}
y_{E}=\beta\sqrt{{\displaystyle \frac{P_{s}L_{c}^{2}}{\left(d_{SR}d_{RE}\right)^{\alpha}}}}\left(\mathbf{h}_{SR}^{\dagger}\mathbf{w}\right)h_{RE}x_{S}+{\displaystyle \frac{\beta h_{RE}n_{R}}{\sqrt{L_{c}d_{RE}^{\alpha}}}}\\
\qquad\qquad\qquad+\sqrt{{\displaystyle \frac{\eta P_{s}L_{c}^{2}}{\left(d_{SJ}d_{JE}\right)^{\alpha}}}}|\mathbf{h}_{SJ}^{\dagger}\mathbf{w}|h_{JE}v+n_{E},
\end{array}\label{eq:received_signal_at_E}
\end{equation}
where $h_{RD}$, $h_{JE}$ and $h_{RE}$ denote the channel coefficients
for the \textbf{R} to \textbf{D}, \textbf{J} to \textbf{E} and\textbf{
R} to \textbf{E} links, respectively, each following complex CSG distribution
with zero mean and unit variance. Substituting (\ref{eq:Beta_eq})
into (\ref{eq:received_signal_at_D}) and (\ref{eq:received_signal_at_E})
and doing some manipulations, the instantaneous signal-to-noise ratios
(SNRs) at \textbf{D}\textit{ $\left(\gamma_{D}\right)$} and\textit{
}\textbf{E}\textit{ }$\left(\gamma_{E}\right)$ are, respectively,
 given as\vspace{-0.1cm}

\begin{equation}
\gamma_{D}={\displaystyle \frac{\gamma_{SR}\gamma_{RD}}{\gamma_{SR}+\gamma_{RD}+1}<\gamma_{D}^{\textrm{ub}}=\frac{\gamma_{SR}\gamma_{RD}}{\gamma_{SR}+\gamma_{RD}}},\label{eq:gamma_d}
\end{equation}
\vspace{-0.3cm}

\begin{equation}
\gamma_{E}={\displaystyle \frac{\gamma_{SR}\gamma_{RE}}{\gamma_{RE}+\left(\gamma_{SR}+1\right)\left(\gamma_{JE}+1\right)}},\label{eq:gamma_e}
\end{equation}
where the superscript \textquoteleft ub' stands for upper-bound, $\gamma_{SR}\triangleq K_{1}|\mathbf{h}_{SR}^{\dagger}\mathbf{w}|^{2},$
$\gamma_{RD}\triangleq K_{2}X_{RD}$, $\gamma_{JE}\triangleq K_{3}|\mathbf{h}_{SJ}^{\dagger}\mathbf{w}|^{2}X_{JE}$,
$\gamma_{RE}\triangleq K_{4}X_{RE}$, $X_{RD}\triangleq|h_{RD}|^{2}$,
$X_{JE}\triangleq|h_{JE}|^{2}$and $X_{RE}\triangleq|h_{RE}|^{2}$.
Furthermore, $K_{1}\triangleq{\displaystyle \left(P_{S}L_{c}d_{SR}^{-\alpha}\right)/\sigma_{n}^{2}}$,
$K_{2}\triangleq{\displaystyle \left(P_{R}L_{c}d_{RD}^{-\alpha}\right)/\sigma_{n}^{2}}$,
$K_{3}\triangleq{\displaystyle \left(\eta P_{S}L_{c}^{2}d_{SJ}d_{JE}\right)^{-\alpha}}/\sigma_{n}^{2}$
and $K_{4}\triangleq\left({\displaystyle P_{R}L_{c}d_{RE}^{-\alpha}}\right)/\sigma_{n}^{2}$.
The exact instantaneous secrecy capacity $\left(C\right)$ is given
as \vspace{-0.2cm}

\begin{equation}
{\displaystyle {\displaystyle C=\frac{1}{2}}\left[C_{D}-C_{E}\right]^{+}},\label{eq:instantaneous_secrecy_capacity}
\end{equation}
where $C_{D}\triangleq\log_{2}\left(1+\gamma_{D}\right)$ , $C_{E}\triangleq\log_{2}\left(1+\gamma_{E}\right)$
and the factor of $1/2$ is due to the fact that both \textbf{R} and
\textbf{J} operate in half-duplex mode. 

\section{Secrecy Analysis }

\subsection{Secure Beamforming}

In this subsection, we present a new secure beamforming scheme and
derive a closed-form solution for the near optimal beamforming vector
assuming a large number of antennas at \textbf{S}. To this end, we
formulate the following optimization problem\vspace{-0.2cm}

\begin{equation}
P_{0}:\quad\arg\max_{\mathbf{w}}\:C\qquad\textrm{s.t.}\;\:||\mathbf{w}||^{2}=1.\label{eq:optimization_prob}
\end{equation}
 The problem $P_{0}$ is non-convex, hence a time-consuming $N$-dimensional
search is required. Motivated by this, in the following we provide
a new closed-form solution of (\ref{eq:optimization_prob}) which
renders not only an efficient evaluation for the near optimal beamforming
vector (i.e., that maximizes an approximation of the objective function)
but also provides a way to characterize a closed-form approximation
of the ergodic secrecy rate (as depicted in subsection 3B).
\begin{prop}
\label{prop:Proposition1}\textup{In the asymptotic large antenna
regime, i.e., as $N\rightarrow\infty$, the near optimal solution
$\left(\mathbf{w}^{o}\right)$ for problem given in (\ref{eq:optimization_prob})
is a linear combination of EB and IB vectors and is given as \vspace{-0.5cm}
}

\begin{equation}
\mathbf{w}^{o}=\overline{t}\mathbf{w}_{I}+\sqrt{1-\overline{t}^{2}}\mathbf{w}_{E},\label{eq:analytical_prop1}
\end{equation}
\textup{where $\mathbf{w}_{I}$ and $\mathbf{w}_{E}$ represent the
IB and EB vectors, respectively, i.e.,} $\mathbf{w}_{I}\triangleq{\displaystyle \mathbf{h}_{SR}/||\mathbf{h}_{SR}||}$,
$\mathbf{w}_{E}\triangleq{\displaystyle \mathbf{h}_{SJ}/||\mathbf{h}_{SJ}||}$,
$\overline{t}\triangleq\sqrt{{\displaystyle \left(-1+\sqrt{1+B_{0}}\right)/B_{0}}}$
\textup{and} $B_{0}\triangleq{\displaystyle N\left(K_{1}\left(\gamma_{RE}/\gamma_{RD}\right)-K_{3}X_{JE}\right)/\left(K_{3}NX_{JE}+\gamma_{RE}\right)}.$ 
\end{prop}
\begin{IEEEproof}
See Appendix A.
\end{IEEEproof}
It is worth pointing out here that Proposition 1 provides some interesting
insights, e.g., when $B_{0}\rightarrow-1$, $\mathbf{w}^{o}\rightarrow\mathbf{w}_{I}$
and $\mathbf{w}^{o}\rightarrow\mathbf{w}_{E}$ for $B_{0}\rightarrow\infty$.
Similarly when $B_{0}\rightarrow0$, $\overline{t}\rightarrow{\displaystyle 1/\sqrt{2}}$
which implies that \textbf{S} achieves a perfect balance between the
IB and EB.

\subsection{Ergodic Secrecy Rate }

From \cite{Zhao2016} and using (\ref{eq:instantaneous_secrecy_capacity}),
$\mathbb{E}\left(C\right)$ provides the ergodic secrecy capacity,
however an exact closed-form expression for $\mathbb{E}\left(C\right)$
appears to be intractable. Nevertheless, we use the following tight
lower bound which is known as ESR ($\bar{C}$) \cite{Zhao2016}\vspace{-0.3cm}

\begin{equation}
\mathbb{E}\left(C\right)\geq\bar{C}=\frac{1}{2}\left[\mathbb{E}\left(C_{D}\right)-\mathbb{E}\left(C_{E}\right)\right]^{+}.\label{eq:bound_ergodic_bound}
\end{equation}
As it clear that the problem of finding a closed-form for (\ref{eq:bound_ergodic_bound})
has been reduced obtaining the closed-form expressions for the $\mathbb{E}\left(C_{D}\right)$
and $\mathbb{E}\left(C_{E}\right)$. However, it is still challenging
to find the closed-form expressions due to the intricate expressions
given in (\ref{eq:gamma_d}) and (\ref{eq:gamma_e}). We present these
closed-form expressions in the following proposition.
\begin{prop}
\label{prop:Proposition_2} \textup{In the asymptotic large antenna
regime, i.e., as $N\rightarrow\infty$ and for given value of \textquoteleft $t$'
the tight closed-form approximations for $\mathbb{E}\left(C_{D}\right)$
and $\mathbb{E}\left(C_{E}\right)$ are given in (\ref{eq:ergodic_Cd_approximation})
and (\ref{eq:ergodic_Ce_approximation}), respectively.}
\end{prop}
\begin{IEEEproof}
See Appendix B.
\end{IEEEproof}

\section{Numerical Examples and Simulations}

In this section, we provide numerical examples with $\eta=0.8$ and
$L_{C}=0.1$. The remaining system parameters are given in the captions
of each figure. To begin with, we investigate the impact of beamforming
vector, $\mathbf{w}$, on the instantaneous secrecy capacity in Fig.
1. It is shown in Fig. 1 that when the channel condition of \textbf{R}
to \textbf{E} link is better than that of the \textbf{R} to \textbf{D}
link, \textbf{S} tends to beamform more energy towards \textbf{J}\textit{
}as compared to beamform information signal towards \textbf{R} to
confound\textbf{ E}. It is interesting to see that the role of AN
at \textbf{J} is very important even when \textbf{R} to \textbf{D}
link is better than the \textbf{R} to \textbf{E} link except in the
extreme case when the channel amplitude of \textbf{R} to \textbf{E}
link approaches zero. In this case, \textbf{S} tends to beamform towards\textbf{
R} instead of \textbf{J}. Moreover, we notice that the secrecy capacity
varies significantly with \textquoteleft t' and its optimal value
is quite close to that obtained through Proposition \ref{prop:Proposition1}.
Furthermore, we investigate the impact of transmission powers $\left(P_{S}\textrm{ and }P_{R}\right)$
on $\overline{t}$ in Fig. 2. We observe that $\overline{t}$ increases
with $P_{R}$, approaching beams of equal strength (i.e., $\overline{t}\rightarrow1/\sqrt{2}$)
and it decreases with $P_{S}$. This observation is also in agreement
with Proposition 1, thus verifies (\ref{eq:analytical_prop1}). 

In Fig. 3, the ergodic secrecy capacity is plotted using Monte Carlo
simulations corresponding to (\ref{eq:instantaneous_secrecy_capacity})
upon averaging over $10^{5}$ channel realizations. It is clear from
Fig. 3 that our derived approximation in Proposition 2 is tight with
these simulations for various values of the transmission powers and
\textquoteleft $t$'. Moreover, Fig. 3 also shows that our proposed
beamforming scheme achieves a higher ESR as compared to the \textquoteleft no-jammer'
case as well as the trivial EB and IB schemes. It is worth pointing
out here that global knowledge of perfect instantaneous channel state
information (CSI) including that of the wiretap links (as considered
in \cite{Zou2013})\textbf{ }is required to calculate \textquoteleft $t$'
in both Fig. 1 and Fig. 2. However in Fig. 3, we took $t=0.35$ by
replacing random variables (RVs) in $B_{0}$ by their respective averages,
thus the value of \textquoteleft $t$' remains fixed for all channel
realizations (therefore only statistical knowledge of CSI for \textbf{R
}to\textbf{ E }and\textbf{ R }to \textbf{D} links is required). Interestingly,
even for a fixed \textquoteleft $t$', the ESR is almost equal to
the case in which \textquoteleft $t$' is selected optimally for each
channel realization. 

\begin{figure}[tb]
\includegraphics[width=3.8in,height=3in]{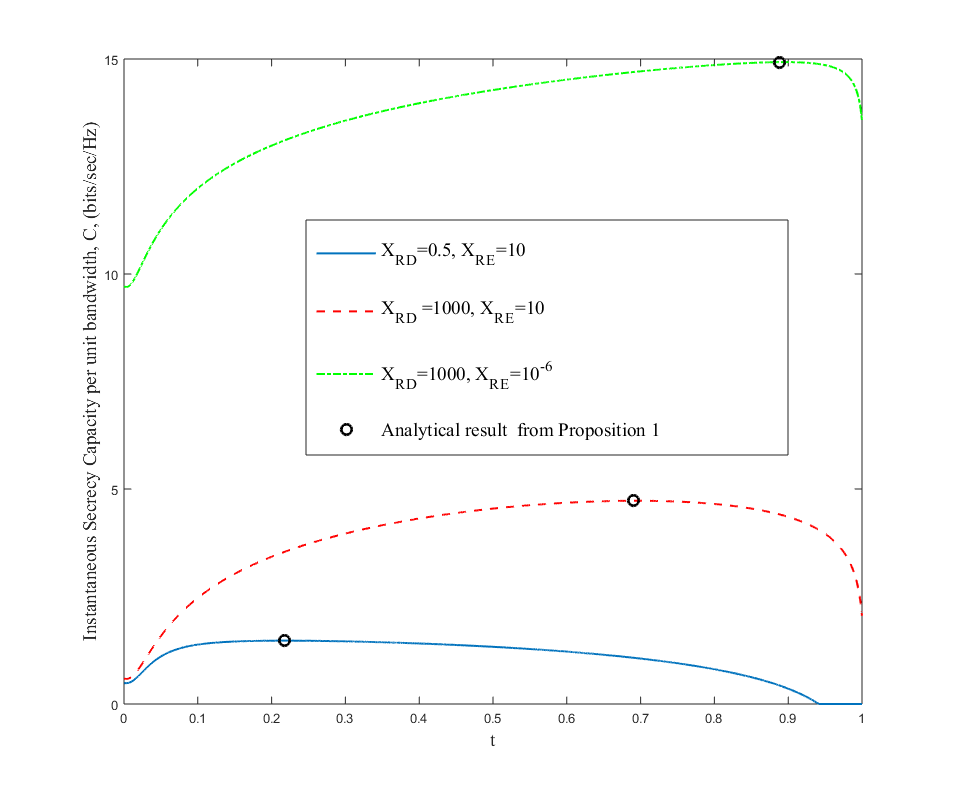}\caption{Plot of instantaneous secrecy capacity using (\ref{eq:instantaneous_secrecy_capacity})
for various values of $t$. $d_{SR}=d_{SJ}=d_{RD}=d_{RE}=d_{JE}=100$m,
$\alpha=3$, $P_{S}=P_{R}=30$ dBm, $\sigma_{n}^{2}=-110$ dBm and
$N=200.$ $X_{JE}=1$, $\mathbf{h}_{SR}$ and $\mathbf{h}_{SJ}$ are
generated in MATLAB in same sequence with rng ($10$). }

\end{figure}

\begin{figure}[tb]

\includegraphics[width=3.8in,height=3in]{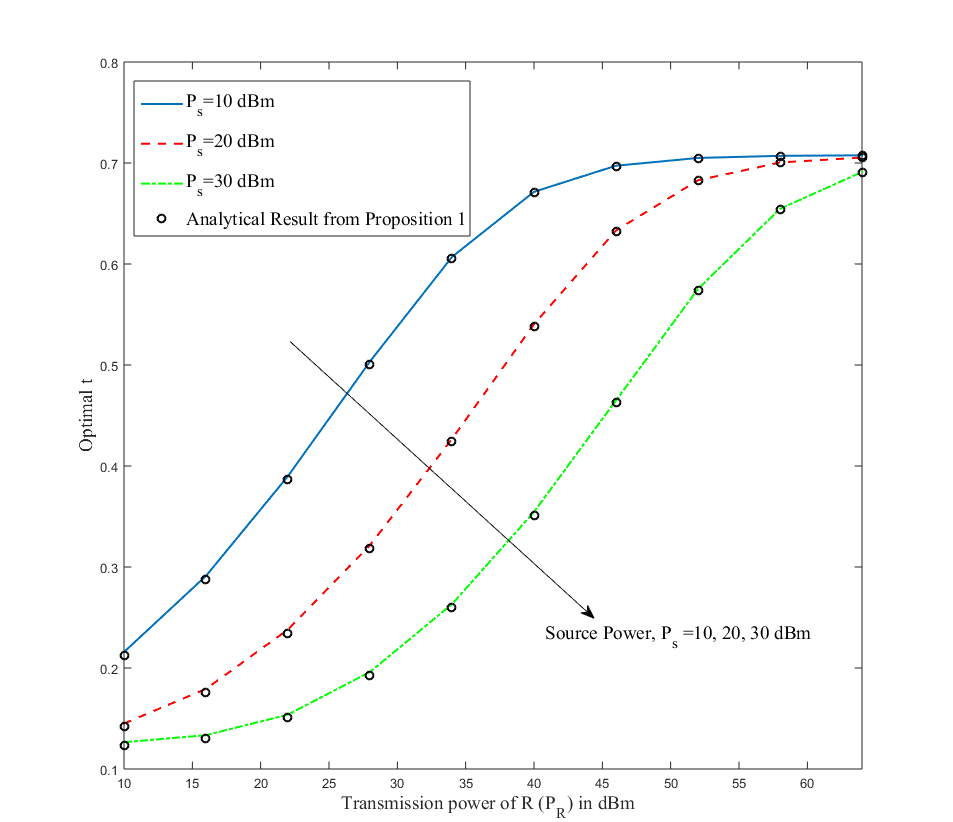}

\caption{Plot of instantaneous secrecy capacity using (\ref{eq:instantaneous_secrecy_capacity})
for various values of $P_{R}$ and $P_{S}$. $d_{SR}=d_{SJ}=d_{RD}=20\textrm{m},\;d_{RE}=30\textrm{m},\;d_{JE}=10$m,
$\alpha=3$, $\sigma_{n}^{2}=-80$ dBm and $N=500.$ $X_{RE}=X_{RD}=1$,
$X_{JE}=0.82,$ $\mathbf{h}_{SR}$ and $\mathbf{h}_{SJ}$ are generated
in MATLAB in same sequence with rng ($100$).}
\end{figure}

\begin{figure}[tbh]
\includegraphics[width=3.8in,height=3in]{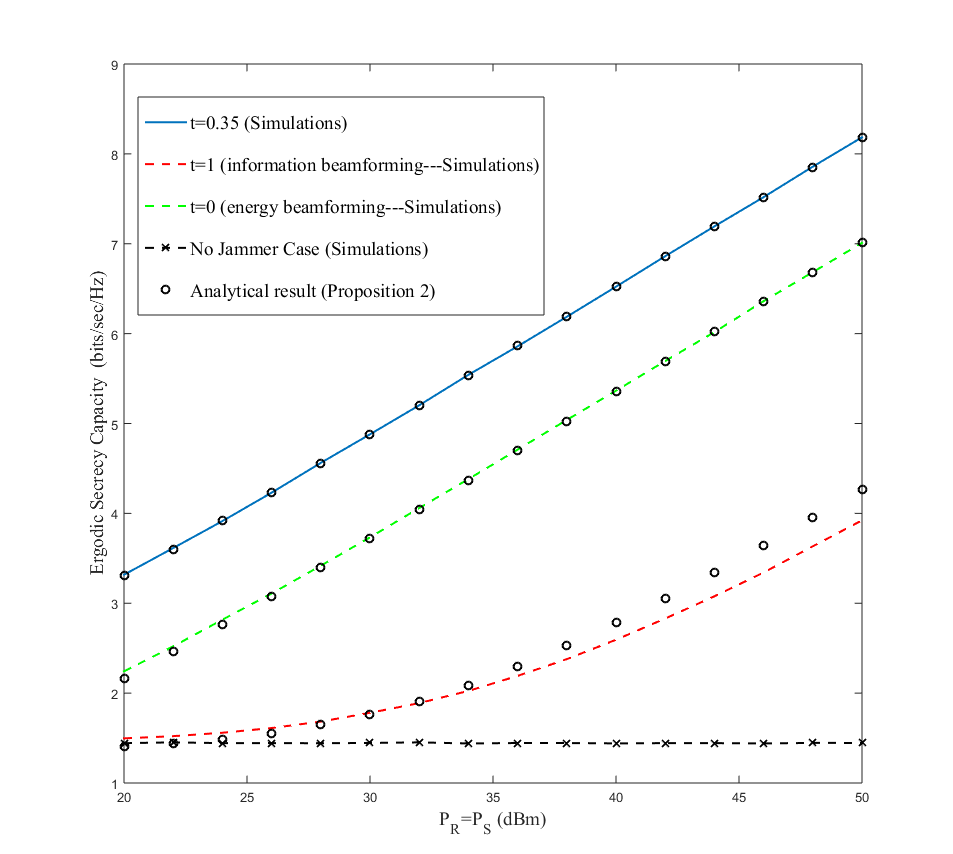}\caption{Plot of ergodic secrecy capacity through simulations and ESR using
(\ref{eq:bound_ergodic_bound}) for various values of $P_{R}$ and
$P_{S}$. $d_{SR}=d_{SJ}=50\textrm{m},\;d_{RD}=30\textrm{m},\;d_{RE}=60\textrm{m,}\;d_{JE}=40$m,
$\alpha=2.7$, $\sigma_{n}^{2}=-80$ dBm and $N=300.$ }

\end{figure}

\section{Conclusion}

In this correspondence, we proposed a secure beamforming scheme which
is shown to achieve a higher ergodic secrecy rate compared to the
trivial information and energy beamforming schemes. Moreover, we present
a new closed-form solution for the proposed beamforming vector which
maximizes the ergodic secrecy capacity. We also derived a new closed-form
approximation for the ergodic secrecy rate with a given beamforming
vector. We observed that, in general, the source tends to transmit
a stronger beam towards the wireless powered jammer as compared to
the relay to maximize the ergodic secrecy rate except for the scenario
in which the relay to eavesdropper channel is extremely weak.

\section*{Appendix A\protect \\
Proof of Proposition \ref{prop:Proposition1}}

Although the objective function of (\ref{eq:optimization_prob}) is
considerably different from that given in \cite{Waqar2019}, $\mathbf{w}$
admits the same generic form\footnote{It is worth mentioning here that only the generic form of $\mathbf{w}$
remains same, our final solution is different from that of \cite{Waqar2019}
due to significant difference in the objective functions. } and for $N\rightarrow\infty,$ we can have \vspace{-0.3cm}

\begin{equation}
\mathbf{w}=t\mathbf{w}_{I}+\sqrt{1-t^{2}}\mathbf{w}_{E},\label{eq:compact_form_w}
\end{equation}
where $t\:\in\:\left[0,1\right]$. Substituting (\ref{eq:compact_form_w})
into the upper bound of (\ref{eq:gamma_d}) we can write \vspace{-0.3cm}

\begin{equation}
\gamma_{D}\backsimeq{\displaystyle \widetilde{\gamma}_{D}\left(t\right)=\frac{\widetilde{\gamma}_{SR}\left(t\right)\gamma_{RD}}{\widetilde{\gamma}_{SR}\left(t\right)+\gamma_{RD}}},\label{eq:gamma_d_approximation}
\end{equation}
where leveraging law of large numbers we put $||\mathbf{h}_{SR}||^{2}\backsimeq N$
and can write $\widetilde{\gamma}_{SR}\left(t\right)\triangleq K_{1}\left[Nt^{2}+\left(1-t^{2}\right)Z_{SR}\right]$
with $Z_{SR}\triangleq|\mathbf{h}_{SR}^{\dagger}\mathbf{w_{E}}|^{2}$
is an exponential RV of unit mean \cite{Waqar2019}. Using a similar
approach as adopted in \cite{Krikidis2009}, (\ref{eq:gamma_e}) can
be upper-bounded as \vspace{-0.3cm}

\begin{equation}
\gamma_{E}<\min\left(\gamma_{SR},{\displaystyle {\displaystyle {\displaystyle \frac{\gamma_{RE}}{\gamma_{JE}}}}}\right).\label{eq:gamma_e_bound}
\end{equation}
Now substitute (\ref{eq:compact_form_w}) into (\ref{eq:gamma_e_bound})
and using again $||\mathbf{h}_{SJ}||^{2}\backsimeq N$, we can write\vspace{-0.3cm}

\begin{equation}
\begin{array}[b]{c}
\gamma_{E}\backsimeq\widetilde{\gamma}_{E}\left(t\right)=\min\left(\widetilde{\gamma}_{SR}\left(t\right),{\displaystyle \frac{\gamma_{RE}}{\widetilde{\gamma}_{JE}\left(t\right)}}\right)\end{array},\label{eq:gamma_e_approximation}
\end{equation}
where $\widetilde{\gamma}_{JE}\left(t\right)\triangleq K_{3}\left[N\left(1-t^{2}\right)+t^{2}Z_{SJ}\right]X_{JE}$
and $Z_{SJ}\triangleq|\mathbf{h}_{SJ}^{\dagger}\mathbf{w_{I}}|^{2}$
is an exponential RV of unit mean. Now with $\widetilde{\gamma}_{D}\left(t\right)\gg1,$
the approximation of (\ref{eq:instantaneous_secrecy_capacity}) can
be written as \vspace{-0.3cm}

\begin{equation}
{\displaystyle {\displaystyle \widetilde{C}\left(t\right)=\frac{1}{2}}\left[\log_{2}\left(\widetilde{\gamma}_{D}\left(t\right)\right)-\log_{2}\left(1+\widetilde{\gamma}_{E}\left(t\right)\right)\right]^{+}}.\label{eq:secrecy_capacity_approximation}
\end{equation}
Furthermore, using (\ref{eq:secrecy_capacity_approximation}) and
after some manipulation, we can reformulate the problem $P_{0}$ as\vspace{-0.3cm}

\begin{equation}
\!\!\!P_{1}:\;\overline{t}=\arg\min_{t}\left[\min\left(g_{1}\left(t\right),g_{2}\left(t\right)\right)\right],\;\textrm{s.t.}\;0\leq t\leq1,\label{eq:P1_problem}
\end{equation}
where $g_{1}\left(t\right)\triangleq{\displaystyle \frac{1+\widetilde{\gamma}_{SR}\left(t\right)}{\widetilde{\gamma}_{D}\left(t\right)}}$
and $g_{2}\left(t\right)\triangleq{\displaystyle {\displaystyle \frac{1}{\widetilde{\gamma}_{D}\left(t\right)}}\left[1+{\displaystyle \frac{\gamma_{RE}}{\widetilde{\gamma}_{JE}\left(t\right)}}\right]}$.
It is interesting to note that for $g_{1}\left(t\right)\leq g_{2}\left(t\right)$
, we have $\widetilde{\gamma}_{D}\left(t\right)\leq\widetilde{\gamma}_{E}\left(t\right)$
and $\widetilde{C}\left(t\right)$ is always equal to zero $\forall t$.
Therefore, problem $P_{1}$ can be simplified as\vspace{-0.2cm}

\begin{equation}
P_{2}:\;\overline{t}=\arg\min_{t}g_{2}\left(t\right),\;\textrm{s.t.}\quad0\leq t\leq1.\label{eq:Problem_P2}
\end{equation}
Now take first derivative of the objective function given in (\ref{eq:Problem_P2})
with respect to \textquoteleft $t$', (note that $Z_{SR}$ and $Z_{SJ}$
related terms in $\widetilde{\gamma}_{SR}\left(t\right)$ and $\widetilde{\gamma}_{JE}\left(t\right)$
can be ignored as $N\rightarrow\infty$), substitute $t^{2}=y$ and
after some manipulation, we get\vspace{-0.2cm}

\begin{equation}
B_{0}y^{2}+2y-1=0,\label{eq:quadratic_equation}
\end{equation}
$B_{0}\triangleq{\displaystyle N\left(K_{1}\left(\gamma_{RE}/\gamma_{RD}\right)-K_{3}X_{JE}\right)/\left(K_{3}NX_{JE}+\gamma_{RE}\right)}$
with $B_{0}\in\left[-1,\infty\right).$ Using the quadratic formula,
we get the following two roots for (\ref{eq:quadratic_equation})\vspace{-0.3cm}

\begin{equation}
y_{1}={\displaystyle \frac{-1-\sqrt{1+B_{0}}}{B_{0}}\quad}\textrm{and}\quad y_{2}={\displaystyle \frac{-1+\sqrt{1+B_{0}}}{B_{0}}.}
\end{equation}
Note that $y_{1}\geq1$ for $-1\leq B_{0}<0$ and $y_{1}<0$ for $B_{0}>0$.
Moreover for $B_{0}\geq-1$, we have $0\leq y_{2}\leq1$, therefore
we get unique minima for $P2$ at $\overline{t}=\sqrt{y_{2}}$ . 

\section*{Appendix B\protect \\
Proof of Proposition \ref{prop:Proposition_2}}

\subsection*{Derivation of $\mathbb{E}\left(C_{D}\right)$: }

By adopting approach as in \cite{Waqar2010}, using (\ref{eq:gamma_d})
and $\widetilde{\gamma}_{SR}\left(t\right)$ from Appendix A, we can
write (\ref{eq:ergodic_Cd_approximation}). Given the fact that for
any RV $X,$ $\mathbb{E}\left(\ln\left\{ 1+X\right\} \right)={\displaystyle \int_{0}^{\infty}\ln\left\{ 1+x\right\} f_{X}\left(x\right)dx,}$
therefore using eq. (2.5.2.1) of \cite{Prudnikov1992}, the identity
$-\textrm{\ensuremath{E_{i}}}\left(-x\right)=\textrm{\ensuremath{E_{1}}}\left(x\right)$
and after some manipulation, we get (\ref{eq:CD1_and_CD2}) and (\ref{eq:CD3_1})

\begin{algorithm*}[tbh]
\begin{equation}
\mathbb{E}\left(C_{D}\right)\backsimeq\frac{1}{\ln\left(2\right)}\left[\underbrace{\mathbb{E}\left(\ln\left\{ 1+\mu_{D}Z_{SR}\right\} \right)}_{C_{D_{1}}}+\underbrace{\mathbb{E}\left(\ln\left\{ 1+K_{2}X_{RD}\right\} \right)}_{C_{D_{2}}}-\underbrace{\mathbb{E}\left(\ln\left\{ \left(1+{\displaystyle \mu_{D}Z_{SR}}+\lambda_{D}X_{RD}\right)\right\} \right)}_{C_{D_{3}}}\right],\label{eq:ergodic_Cd_approximation}
\end{equation}

where ${\displaystyle \mu_{D}\triangleq\frac{K_{1}\left(1-t^{2}\right)}{1+K_{1}Nt^{2}}}$
and ${\displaystyle \lambda_{D}\triangleq\frac{K_{2}}{1+K_{1}Nt^{2}}}.$
Moreover, 

\begin{equation}
C_{D_{1}=}\exp\left({\displaystyle \frac{1}{\mu_{D}}}\right)\textrm{\ensuremath{E_{1}}}\left({\displaystyle \frac{1}{\mu_{D}}}\right)\quad\textrm{and}\quad C_{D_{2}}=\exp\left({\displaystyle \frac{1}{K_{2}}}\right)\textrm{\ensuremath{E_{1}}}\left({\displaystyle \frac{1}{K_{2}}}\right).\label{eq:CD1_and_CD2}
\end{equation}

\begin{equation}
C_{D_{3}}={\displaystyle \frac{1}{\lambda_{D}-\mu_{D}}\left[\lambda_{D}\exp\left({\displaystyle \frac{1}{\lambda_{D}}}\right)\textrm{\ensuremath{E_{1}}}\left({\displaystyle \frac{1}{\lambda_{D}}}\right)-\mu_{D}\exp\left({\displaystyle \frac{1}{\mu_{D}}}\right)\textrm{\ensuremath{E_{1}}}\left({\displaystyle \frac{1}{\mu_{D}}}\right)\right]\qquad\textrm{for \ensuremath{\lambda_{D}\neq\mu_{D}}}.}\label{eq:CD3_1}
\end{equation}
\end{algorithm*}

\begin{algorithm*}[tbh]
\begin{equation}
\mathbb{E}\left(C_{E}\right)\backsimeq\frac{1}{\ln\left(2\right)}\left[\underbrace{\mathbb{E}\left(\ln\left\{ 1+K_{4}X_{RE}+\mu_{E}X_{JE}\right\} \right)}_{C_{E_{1}}}-\underbrace{\mathbb{E}\left(\ln\left\{ \left(1+\mu_{E}X_{JE}+{\displaystyle \lambda_{E}X_{RE}}\right)\right\} \right)}_{C_{E_{2}}}\right],\label{eq:ergodic_Ce_approximation}
\end{equation}

where $\mu_{E}\triangleq K_{3}\left[N\left(1-t^{2}\right)+t^{2}\right]$
and $\lambda_{E}\triangleq{\displaystyle \frac{K_{4}}{1+K_{1}\left[Nt^{2}+\left(1-t^{2}\right)\right]}}$.
Moreover, 

\begin{equation}
C_{E_{1}}={\displaystyle \frac{1}{\mu_{E}-K_{4}}\left[\mu_{E}\exp\left({\displaystyle \frac{1}{\mu_{E}}}\right)\textrm{\ensuremath{E_{1}}}\left({\displaystyle \frac{1}{\mu_{E}}}\right)-K_{4}\exp\left({\displaystyle \frac{1}{K_{4}}}\right)\textrm{\ensuremath{E_{1}}}\left({\displaystyle \frac{1}{K_{4}}}\right)\right]\qquad\textrm{for \ensuremath{K_{4}\neq\mu_{E}}}.}\label{eq:CE11}
\end{equation}

\begin{equation}
C_{E_{2}}={\displaystyle {\displaystyle \frac{1}{\lambda_{E}-\mu_{E}}\left[\lambda_{E}\exp\left({\displaystyle \frac{1}{\lambda_{E}}}\right)\textrm{\ensuremath{E_{1}}}\left({\displaystyle \frac{1}{\lambda_{E}}}\right)-\mu_{E}\exp\left({\displaystyle \frac{1}{\mu_{E}}}\right)\textrm{\ensuremath{E_{1}}}\left({\displaystyle \frac{1}{\mu_{E}}}\right)\right]\qquad\textrm{for \ensuremath{\lambda_{E}\neq\mu_{E}}}.}}\label{eq:CE21}
\end{equation}
\end{algorithm*}

In case for $\ensuremath{\lambda_{D}=\mu_{D}}$, using eq. (2.5.2.11)
of \cite{Prudnikov1992} we can replace (\ref{eq:CD3_1}) with \vspace{-0.4cm}

\begin{equation}
C_{D_{3}}=G_{3,2}^{1,3}\left[\mu_{D}{\displaystyle {\displaystyle \left|\begin{array}{c}
-1,\,1,\,1\\
1,\,0
\end{array}\right.}}\right].\label{eq:CD3_2}
\end{equation}
Now placing (\ref{eq:CD1_and_CD2}) and (\ref{eq:CD3_1}) or (\ref{eq:CD3_2})
into (\ref{eq:ergodic_Cd_approximation}), we get the desired result.

\subsection*{Derivation of $\mathbb{E}\left(C_{E}\right)$: }

Again adopting the approach of \cite{Waqar2010} and after few manipulation,
we can write (\ref{eq:ergodic_Ce_approximation}). The closed-form
expression for (\ref{eq:ergodic_Ce_approximation}) appears to be
intractable, thus we replaced RVs $Z_{SR}=Z_{SJ}$ with their respective
means (which are unity). As the remaining derivation for $\mathbb{E}\left(C_{E}\right)$
is similar to that of $\mathbb{E}\left(C_{D}\right)$, for sake of
brevity we skipped the steps. Furthermore, for $\ensuremath{K_{4}=\mu_{E}}$
or $\ensuremath{\lambda_{E}=\mu_{E}}$, we can replace (\ref{eq:CE11})
or (\ref{eq:CE21}) accordingly with 

\begin{equation}
C_{E_{1}}\left(\textrm{or}\;C_{E_{2}}\right)=G_{3,2}^{1,3}\left[\mu_{E}{\displaystyle {\displaystyle \left|\begin{array}{c}
-1,\,1,\,1\\
1,\,0
\end{array}\right.}}\right].
\end{equation}
Now the evaluation for $\mathbb{E}\left(C_{E}\right)$ can be done
using (\ref{eq:ergodic_Ce_approximation}).

\bibliographystyle{IEEEtran}
\bibliography{EH_Jammer_Paper_Final}

\end{document}